\documentclass[fleqn,10pt]{wlscirep}

\usepackage{amsmath}
\usepackage{graphicx}
\usepackage{dcolumn}
\usepackage{bm}
\usepackage{float}

\title{Direct acceleration of electrons by a CO$_{2}$ laser in a curved plasma waveguide}

\author[1,2,*]{Longqing Yi}
\author[1]{Alexander Pukhov}
\author[2,3]{Baifei Shen}
\affil[1]{Institut fuer Theoretische Physik I, Heinrich-Heine-Universitaet Duesseldorf, Duesseldorf, 40225 Germany}
\affil[2]{State Key Laboratory of High Field Laser Physics, Shanghai Institute of Optics and Fine Mechanics, Chinese Academy of Sciences, P.O. Box 800-211, Shanghai 201800, China}
\affil[3]{Collaborative Innovation Center of IFSA (CICIFSA),Shanghai Jiao Tong University, Shanghai 200240, China}

\affil[*]{corresponding author: yi@uni-duesseldorf.de}

\begin{abstract}

\textbf{Laser plasma interaction with micro-engineered targets at relativistic
intensities has been greatly promoted by recent progress in the high
contrast lasers and the manufacture of advanced micro- and nano-structures.
This opens new possibilities for the physics of laser-matter interaction.
Here we propose a novel approach that leverages the advantages of
high-pressure CO$_{2}$ laser, laser-waveguide interaction, as well
as micro-engineered plasma structure to accelerate electrons to peak
energy greater than 1 GeV with narrow slice energy spread ($\sim1\%$)
and high overall efficiency. The acceleration gradient is 26 GV/m
for a 1.3 TW CO$_{2}$ laser system. The micro-bunching of a long
electron beam leads to the generation of a chain of ultrashort electron
bunches with the duration roughly equal to half-laser-cycle. These
results open a way for developing a compact and economic electron
source for diverse applications.}
\end{abstract}

\begin{document}

\flushbottom
\maketitle
%
%
\thispagestyle{empty}

\section*{Introduction}

Laser wakefield acceleration \cite{s1} of particles to relativistic energies
has been greatly promoted by the invention of chirped-pulse-amplification
(CPA) \cite{s2} more than twenty years ago. The ultrashort laser
pulses with huge peak powers enabled by the CPA technique allowed
for quasimonoenergetic electron bunches to be generated in underdense
plasmas in so-called bubble regime \cite{bubble,s3,s4,s5} of laser
wakefield. Pioneering experiments had reported that electrons with
a few percent energy spread and sub-milliradian divergences beyond
4.2 GeV can be produced \cite{s6}, which demonstrates the impressive
progress in plasma-based acceleration. However, a significant drawback
is the traditional CPA lasers use TiSa crystals which deliver an average
power of a few Watts only at a low overall efficiency \cite{s7}.
On the other hand, well-known for its industrial applications, the
overall efficiency (5-20\% from wall plug) of CO$_{2}$ laser is among
the highest of all lasers. Hence, it is the most economic choice when
considering high energy physics applications, where high luminosities
are usually required.

Nowadays, high-pressure CO$_{2}$ laser has already reached multi-Terawatt-level
\cite{s8} and been successfully applied for a series of proton acceleration
experiments \cite{s9,s10}. However, it has been less progresses in
CO$_{2}$ laser-driven wakefield acceleration \cite{s11,s12}, mainly
because of the difficulty with building ultra-short CO$_{2}$ laser
system. In general, it is well known that the longitudinal dimension
of the driver of plasma wakefield should be comparable to plasma wavelength
in order to resonantly excite a bubble \cite{bubble,s13,s14}, such that
for the typical CO$_{2}$ laser pulse duration ($\tau\sim10$ ps),
an extremely low-density plasma ($n_{e}\sim10^{13}$ cm$^{-3}$) is
required, which is of little interest to the accelerator community
since the maximum acceleration gradient (i.e. wave breaking field)
is on the same order of magnitude with the conventional RF accelerators.

\begin{figure}[!b]
\centering
\includegraphics[width=15.5cm]{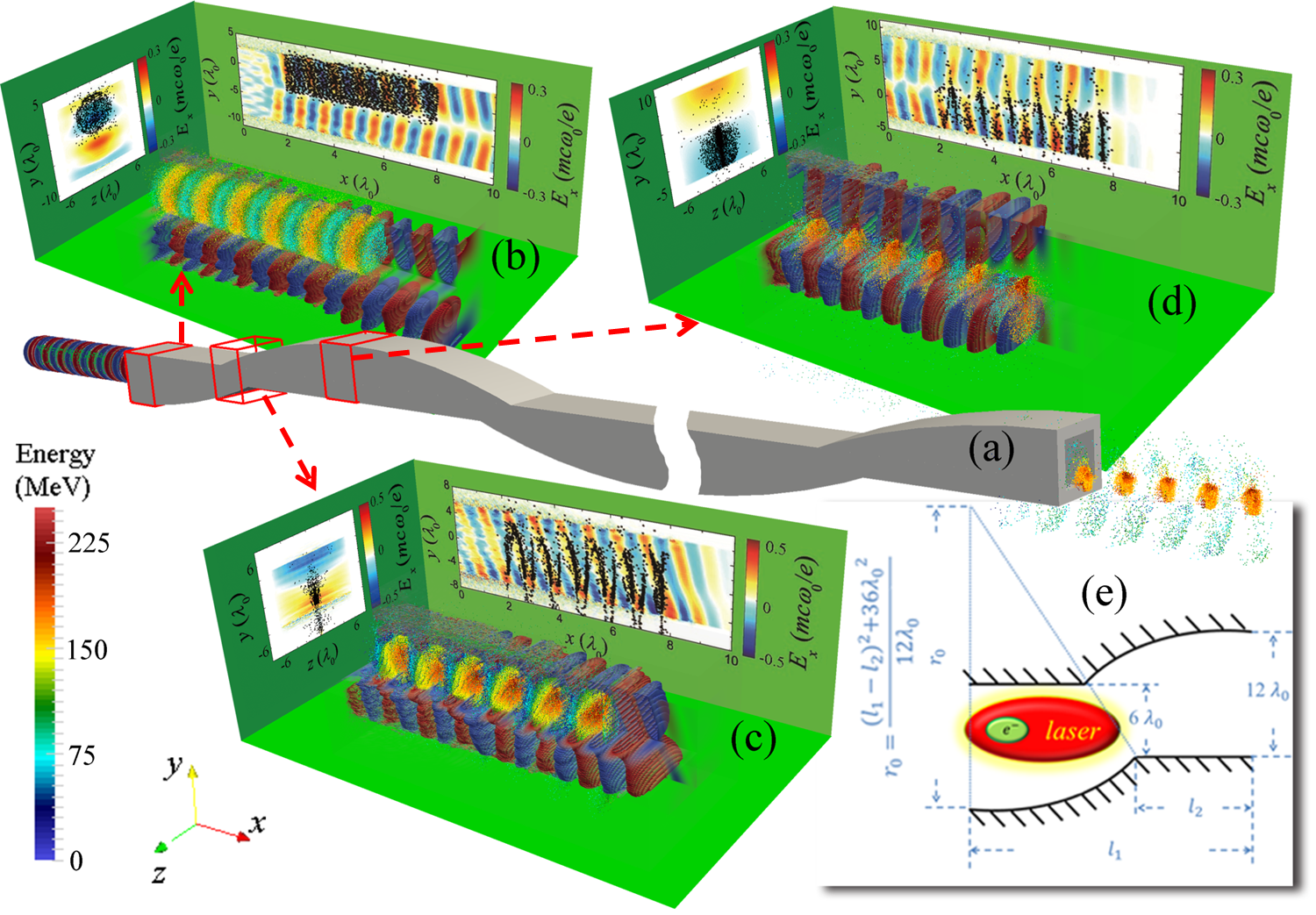}
\caption{(Color online.) (a) Sketch of direct laser acceleration
of electron in a CPW. A linearly polarized CO$_{2}$ laser and a relativistic
electron beam are injected into a CPW from the left side. The electrons
located in the right phase can be accelerated by the TM modes, resulting
in the generation of a chain of energetic ultrashort electron bunches
(right exit of CPW). (b-d) Shows the longitudinal electric field and
electron motion at different propagating distances (marked by the
red cube-frame in (a)), where a quarter of $E_{x}$ field are removed
to avoid overlapping, and the electron energy is presented by the
color. The figure on the back and left walls in (b-d) presents the
$E_{x}$ field and electron position (black dots) at longitudinal
slice $z=0$ and transverse cross-section $x=6.5\lambda_{0}$, respectively.
The waveguide is properly designed with a curvature in the polarization
direction, detail dimension of half of the CPW period is shown in
(e) for x-y plain (the waveguide is uniform along z-direction, and
the size in z dimension is $12\lambda_{0}$ in the presented 3D PIC
simulation). The $l_{1}$ is the half-CPW curvature period, and the
ratio of $l_{1}$ and $l_{2}$ is optimized according to the simulation
($l_{2}=0.442l_{1}$). If $l_{1}$ matches the dephasing length, the
electron bunch can continuously gain energy until it overtakes the
entire laser pulse.}
\label{fig:1}
\end{figure}

In parallel, direct laser acceleration (DLA) offers an attractive
alternative \cite{s15,s16}, where no threshold intensity \cite{s17}
and no limitiations on the pulse duration. Normally, a waveguide can
be used to guide laser pulses over distances much larger than the
Rayleigh length $Z_{R}=\pi w_{0}^{2}/\lambda_{0}$ ($w_{0}$ is the
spot size and $\lambda_{0}$ is the laser wavelength), while simultaneously,
transverse magnetic (TM) optical modes are excited in the channel.
The co-propagating electrons in a proper phase can be accelerated
with a peak longitudinal electric field that can be estimated
by \cite{s18,s19,s20}
\begin{align}
\vspace{-10pt}\begin{split}E[GV/cm]\approx\frac{8a_{0}}{R[\mu m]},\end{split}
\vspace{-10pt}
\end{align}
where $a_{0}=eE_{0}/m_{e}c\omega_{0}$ is the normalized laser amplitude, $c$ is the
light velocity in vacuum $m_{e}$ is the mass of an electron,
$e$ is the unit charge, $\omega_{0}$ is the frequency of laser, and $R$ is the radius
of waveguide in $\mu m$. For a 1.3 TW CO$_{2}$ laser pulse with
$a_{0}=5$ and channel radius $R=6\lambda_{0}\approx64$ $\mu m$, the
peak acceleration gradient is roughly 0.64 GV/cm, which compare favorably
to laser wakefield acceleration with ultra-short laser pulses at the
same power level.

However, the slippage between laser phase velocity and the electron
velocity (essentially $c$) forbids the electrons to stay in the acceleration
phase, which sets a limit on the maximum energy that can be aquired.
Historically, periodical grating surfaces \cite{s21,s22} and neutral
gas filling \cite{s23} have been proposed to solve the dephasing
problem, but none of them can survive the relativistic intensities
required for high acceleration gradient even for the pulse duration.
In fully ionized plasma channel, the phase velocity of laser is superluminal.
A corrugated plasma waveguide has been proposed as ultrahigh intensity
optical slow-wave structure, where net energy gain can be achieved
using a radially polarized laser propagating in a density-modulated
gas jet \cite{s17}. More recently, owing to the advancements in laser
pulse cleaning techniques \cite{s24,s25} and 3D direct laser writing
(DLW) of materials \cite{s26}, laser interaction with fine plasma
structure is drawing more and more attention. The micro- and nano-structured
plasma targets have been introduced to manipulate laser matter interactions
\cite{s20,s27,s28,s29} Simulations suggest that the longitudinal
electric field in excess of 1 TV/m can be achieved in an overdense
micro-plasma-waveguide \cite{s20}. An taylored plasma microstructure
that can overcome the phase slippage would allow for enormous acceleration
gradients.

In this letter, we propose a novel electron acceleration scheme using
DLA in a curved plasma channel (CPW) that is capable of generating
energetic ($>$ 1 GeV) ultra-short (duration $\sim$ half-laser-cycle)
electron bunch chain with slice energy spread $\sim$ 1$\%$. These
high-quality electron beams can be widely applied in high energy physics,
study of atomic and molecular dynamics and generating coherent x-rays.
In the presented study, A CPW is used to overcome the phase slippage
as shown in Fig.~1. Inspired by the fact that the longitudinal electric
field in the CPW is anti-symmetric with respect to the propagation
axis in the polarization direction (as shown in Fig.~1(b-d)), we
properly design the spatial period of CPW to match the dephasing length
of electrons. So that the relative displacement between electrons
and laser field in the longitudinal and transverse directions keeps
the electrons in the accelerating phase (for most of the time), enabling
continuous energy gain of witness beam until it overtakes the entire
laser pulse. In addition, a linearly polarized CO$_{2}$ laser beam
is employed as the driver, not only due to the high overall efficiency
discussed above, but also because a long infrared wavelength enables
a large acceleration bracket, which increases the number of particles
per bunch. Nevertheless, the wavelength of drive laser is not mandatory,
it should be carefully chosen according to the applications.

\section*{Results}

\subsection*{PIC simulation results}

Here we first demonstrate the acceleration process with three dimensional (3D)
particle-in-cell (PIC) simulations using the code VLPL \cite{s30}, the parameters
can be found in \textbf{Methods}.The CPW is constituted with an overdense slab and an arc,
the detailed dimensions for the structure are shown in Fig.~1(e).
The initial energy of the injected electron bunch is 100 MeV.
Figures~1(b-d) show the relative motion of the electron bunch and longitudinal
electric field in half of the CPW period, which indicate the transverse
motion of the guided laser beam perfectly compensates the phase slippage
effect. As soon as the longitudinal phase slippage reaches half of
laser cycle, electrons exactly fall into the acceleration phase on
the opposite side of channel as expected. Moreover, since only the
electrons with proper phase ($E_{x}<0$) can be captured and accelerated
within a long electron bunch, a chain of ultrashort electron bunches
is generated. The duration of a single electron bunch is governed
by the laser wavelength, an attosecond electron train can be generated
if a short-wavelength laser is employed as the driver, which may be
applied in the ultrafast electron diffraction and 4D microscopy \cite{s31,s32,s33}.

\begin{figure}[!b]
\centering
\includegraphics[width=13.5cm]{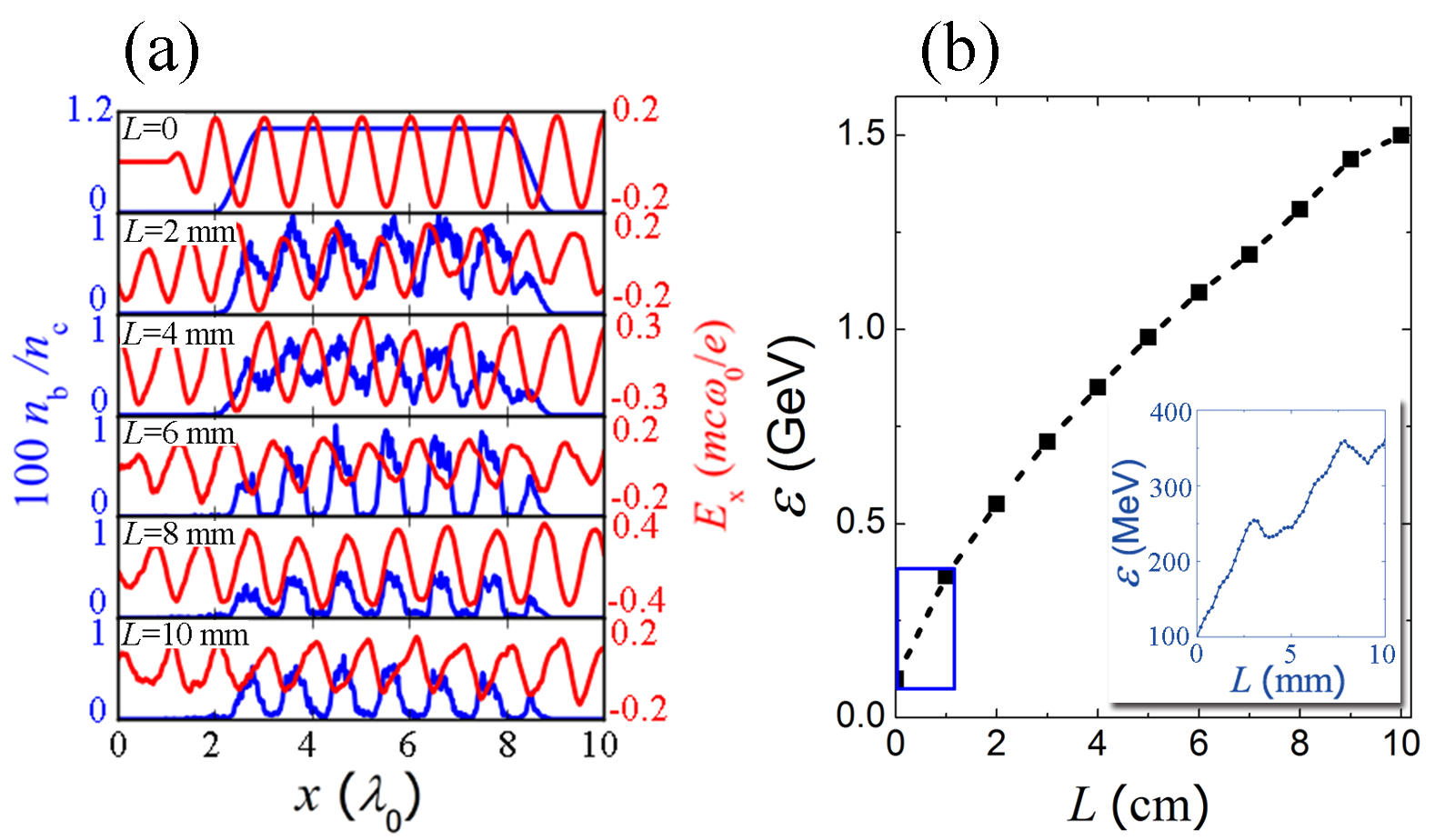}
\caption{(Color online.) (a) The on-axis longitudinal electric
field and witness electron density profile in the first CPW period
at propagation distance $L$ = 0, 2, 4, 6, 8, and 10 mm (b) peak energy
of electron bunch at the end of each CPW period, the inset shows detail
energy evolution in the first CPW period (blue rectangular in (b)).}
\label{fig:2}
\end{figure}

In order to obtain a deeper insight in the laser pulse propagation
and electron acceleration, we perform 2D PIC simulation on the electron
acceleration over 10 centimeters (10 CPW periods).  A long laser
pulse (duration $\sim$ 1 ps) is employed to enable the acceleration
over a long distance. The on-axis longitudinal electric field and witness
electron density profile are plotted in Fig.~2(a) for one CPW period.
One can see that the micro-bunching occurs simultaneously with the
acceleration during the first 2-mm propagation in the CPW. The generated
micro bunches stay in the acceleration phase for most of time.

Figure~2(b) presents the peak energy of electron bunch at the end
of each CPW period, where the final electron energy attained is 1.5
GeV. The acceleration gradient decreases as the laser energy depletes
in the CPW. After 10 cm propagation, the laser loses $80\%$ of its
energy. For electrons with divergence smaller than 1 mrad, the overall
laser-to-electron energy efficiency is roughly 11$\%$. The inset
of Fig.~2(b) shows the peak electron energy evolution in the first
CPW period. The slight energy decrease observed at $L\approx$ 3 mm
and 8 mm is due to the brief passage of the witness bunch through
the decelerating phase as the laser pulse is guided across the beam
axis (see Fig.~1(c) and Fig.~2(a)). The net energy gain in the first
CPW period (1 cm) is 260 MeV that results in an average acceleration
gradient of 26 GV/m, or roughly 40$\%$ of the peak electric field
predicted in Eq.~(1).

\subsection*{Theoretical analysis}

In the following, we try to give an estimation on the electromagnetic
field and acquire basic results for dephasing distance by investigating
laser propagation in a plane plasma waveguide. Further PIC simulation
should be relied on to obtain optimum parameters for a real CPW. Considering
a $y$-polarized laser pulse entering the plasma waveguide along $x$-axis,
the waveguide has a rectangular cross-section in $y$-$z$ plane ($y=-y_{0}\sim y_{0}$,
and $z=-z_{0}\sim z_{0}$). Following the methods in Ref. \cite{s19,s20},
one can easily write the electric and magnetic fields in terms of
two Hertz potentials $\Pi^{e}$ and $\Pi^{h}$ in Cartesian coordinate
system:
\begin{align}
\vspace{-10pt}\begin{split}E_{x}=\frac{\partial^{2}\Pi^{e}}{\partial x^{2}}+k^{2}\Pi^{e},H_{x}=\frac{\partial^{2}\Pi^{h}}{\partial x^{2}}+k^{2}\Pi^{h},\end{split}
\vspace{-10pt}
\end{align}
\begin{align}
\vspace{-10pt}\begin{split}E_{y}=\frac{\partial^{2}\Pi^{e}}{\partial x\partial y}-i\omega\mu\frac{\partial\Pi^{h}}{\partial z},H_{y}=\frac{\partial^{2}\Pi^{h}}{\partial x\partial y}+i\omega\epsilon\frac{\partial\Pi^{e}}{\partial z},\end{split}
\vspace{-10pt}
\end{align}
\begin{align}
\vspace{-10pt}\begin{split}E_{z}=\frac{\partial^{2}\Pi^{e}}{\partial x\partial z}+i\omega\mu\frac{\partial\Pi^{h}}{\partial y},H_{z}=\frac{\partial^{2}\Pi^{h}}{\partial x\partial z}-i\omega\epsilon\frac{\partial\Pi^{e}}{\partial y},\end{split}
\vspace{-10pt}
\end{align}
where
\begin{align}
\vspace{-10pt}\begin{split}\Pi^{e}=\left\{ \begin{aligned}A^{e}\sin(k_{y}y)\cos(k_{z}z)e^{-ik_{x}x},~(|y|<y_{0},|z|<z_{0})\\
B^{e}\cos(k_{z}z)e^{-ik_{x}x-ik_{yp}|y|},~(|y|\geq y_{0},|z|<z_{0})\\
C^{e}\sin(k_{y}y)e^{-ik_{x}x-ik_{zp}|z|},~(|y|<y_{0},|z|\geq z_{0})
\end{aligned}
\right.\end{split}
\vspace{-10pt}
\end{align}
\begin{align}
\vspace{-10pt}\begin{split}\Pi^{h}=\left\{ \begin{aligned}A^{h}\cos(k_{y}y)\sin(k_{z}z)e^{-ik_{x}x},~(|y|<y_{0},|z|<z_{0})\\
B^{h}\sin(k_{z}z)e^{-ik_{x}x-ik_{yp}|y|},~(|y|\geq y_{0},|z|<z_{0})\\
C^{h}\cos(k_{y}y)e^{-ik_{x}x-ik_{zp}|z|},~(|y|<y_{0},|z|\geq z_{0})
\end{aligned}
\right.\end{split}
\vspace{-10pt}
\end{align}

Here $A^{e}$, $B^{e}$... etc are coefficients determined by the
incident laser amplitude, $k=\sqrt{k_{x}^{2}+k_{y}^{2}+k_{z}^{2}}$
is the total wave number in the vacuum core, and $k_{x}$, $k_{y}$,
$k_{z}$ are the wave numbers in each direction, $k_{yp},k_{zp}$
are the transverse wave numbers inside the plasma channel walls. Apparently,
since laser is properly guided in the $x$ direction, $k_{yp},k_{zp}$
are both imaginary.

\begin{figure}[ht]
\centering
\includegraphics[width=\linewidth]{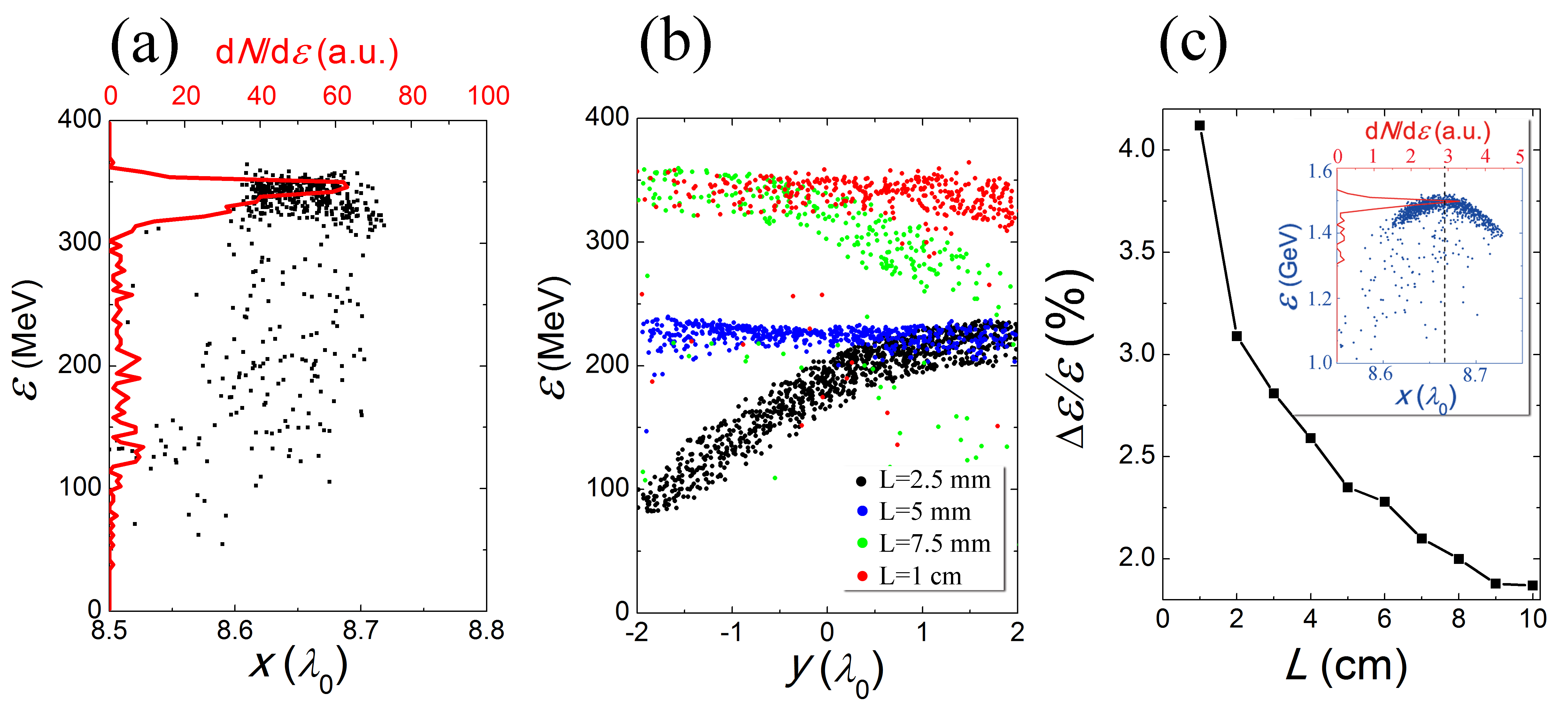}
\caption{((Color online.) (a) Longitudinal and (b) transverse phase
space of electrons in the first CPW period. (c) The relative energy
spread evolution during the 10-cm acceleration. The red line in (a)
shows the electron energy spectrum, the different colors in (b) corresponding
to the phase space at corresponding distance, and the inset of (c)
presents the phase space of electrons at highest energies.}
\label{fig:3}
\end{figure}

At boundaries $y=\pm y_{0}$ and $z=\pm z_{0}$, the tangential components
of both $E$ and $H$ should be continuous, which leads to the following
eigenvalue equations,
\begin{align}
\vspace{-10pt}\begin{split}[(1-\frac{\omega_{p}^{2}}{\omega^{2}})Y_{p}\sin(Y)-Y\cos(Y)][Y_{p}\cos(Y)-Y\sin(Y)]=0,\end{split}
\vspace{-10pt}
\end{align}
\begin{align}
\vspace{-10pt}\begin{split}[(1-\frac{\omega_{p}^{2}}{\omega^{2}})Z_{p}\cos(Z)-Z\sin(Z)][Z_{p}\sin(Z)+Z\cos(Z)]=0,\end{split}
\vspace{-10pt}
\end{align}
where $\omega_{p}$ is the plasma frequency, and $Y=k_{y}y_{0}$,
$Y_{p}=-ik_{yp}y_{0}$, $Z=k_{z}z_{0}$, $Z_{p}=-ik_{zp}z_{0}$, which
satisfy $Y^{2}+Y_{p}^{2}=\frac{\omega_{p}^{2}}{\omega^{2}}k^{2}y_{0}^{2}$,
and $Z^{2}+Z_{p}^{2}=\frac{\omega_{p}^{2}}{\omega^{2}}k^{2}z_{0}^{2}$.

In 2D limit, i.e. $z_{0} \gg y_{0}$, the wave number $k_{z}$ is negligible.
Let $\alpha$ be the 1st root for $Y$ in Eq.~(7) ($\alpha=1.54$),
which corresponding to the lowest TM mode in the waveguide. One can
write the longitudinal electric field according to Eq.~(2) as
\begin{align}
\vspace{-10pt}\begin{split}E_{x}\approx\frac{\alpha m_{0}c^{2}}{y_{0}e}a_{0}\sin(\frac{\alpha y}{y_{0}})e^{-ik_{x}x}.\end{split}
\vspace{-10pt}
\end{align}

\section*{Discussion}

Apparently, the acceleration gradient is only uniform along the $z-$direction,
the nonuniformity in x- and y-direction tends to broaden energy spread
of witness beam during the acceleration. However, the quality of acceleration
can be understood when a short witness electron bunch
is injected into the CPW. The witness bunch stays monoenergetic as
long as it is injected to the position corresponding to the peak acceleration
field. The sufficiently short bunch length ($\sim$1 $\mu m$, occupies
about 10$\%$ of the laser cycle) makes sure it gains little longitudinal energy
chirp as shown in Fig.~3(a). Also, the transverse phase space map
(Fig.~3(b)) illustrates that although the accelerating field varies
along the y-axis, the integration in one period of CPW is uniform.
So the electrons at different transverse position gains the same amount
of energy in one CPW period. As a result, when a short (1 $\mu m$)
electron bunch is injected into the proper accelerating phase, a quasi-monoenergetic
electron beam can be obtained as shown in Fig.~3(c). The relative
r.m.s energy spread at highest energy (1.5 GeV) is about 2$\%$
for the whole beam, this result can be further optimized by employing shorter
witness bunch. The absolute energy spread $\triangle\varepsilon$
increases slightly owing to longitudinal energy chirp as shown by
the inset phase space map, and the r.m.s slice energy spread is 0.83$\%$.

The above results also allow us to derive the matching condition which
is crucial to the proposed scheme. In a sufficiently short propagation
distance $dx$, the change in transverse size of CPW is negligible,
the CPW can be treated as a plane waveguide, and the phase slippage
between relativistic electrons ($v\approx c$) and TM$_{10}$ mode
is $\Delta(x)dx=\alpha^{2}\lambda_{0}^{2}/2\pi^{2}h(x)^{2}dx$, where
$h(x)$ is the CPW dimension along $y$-direction. The matching condition
states that the phase slippage in half CPW period must be equal to
$\lambda_{0}/2$, i.e.

\begin{align}
\vspace{-10pt}\begin{split}\int_{0}^{l_{1}}\Delta(x)dx=2\int_{0}^{l_{2}}\frac{\alpha^{2}\lambda_{0}^{2}}{2\pi^{2}h_{1}(x)^{2}}dx+\int_{l_{2}}^{l_{1}-l_{2}}\frac{\alpha^{2}\lambda_{0}^{2}}{2\pi^{2}h_{2}(x)^{2}}dx=\frac{\lambda_{0}}{2},
\end{split}
\vspace{-10pt}
\end{align}
where $h_{1}(x)=12\lambda_{0}-r_{0}+\sqrt{r_{0}^{2}-x^{2}}$ and $h_{2}(x)=24\lambda_{0}-2r_{0}+\sqrt{r_{0}^{2}-x^{2}}+\sqrt{r_{0}^{2}-(x-l_{1})^{2}}$
are the expressions for $h(x)$ in the flat and curve areas, respectively.

\begin{figure}[ht]
\centering
\includegraphics[width= 10cm]{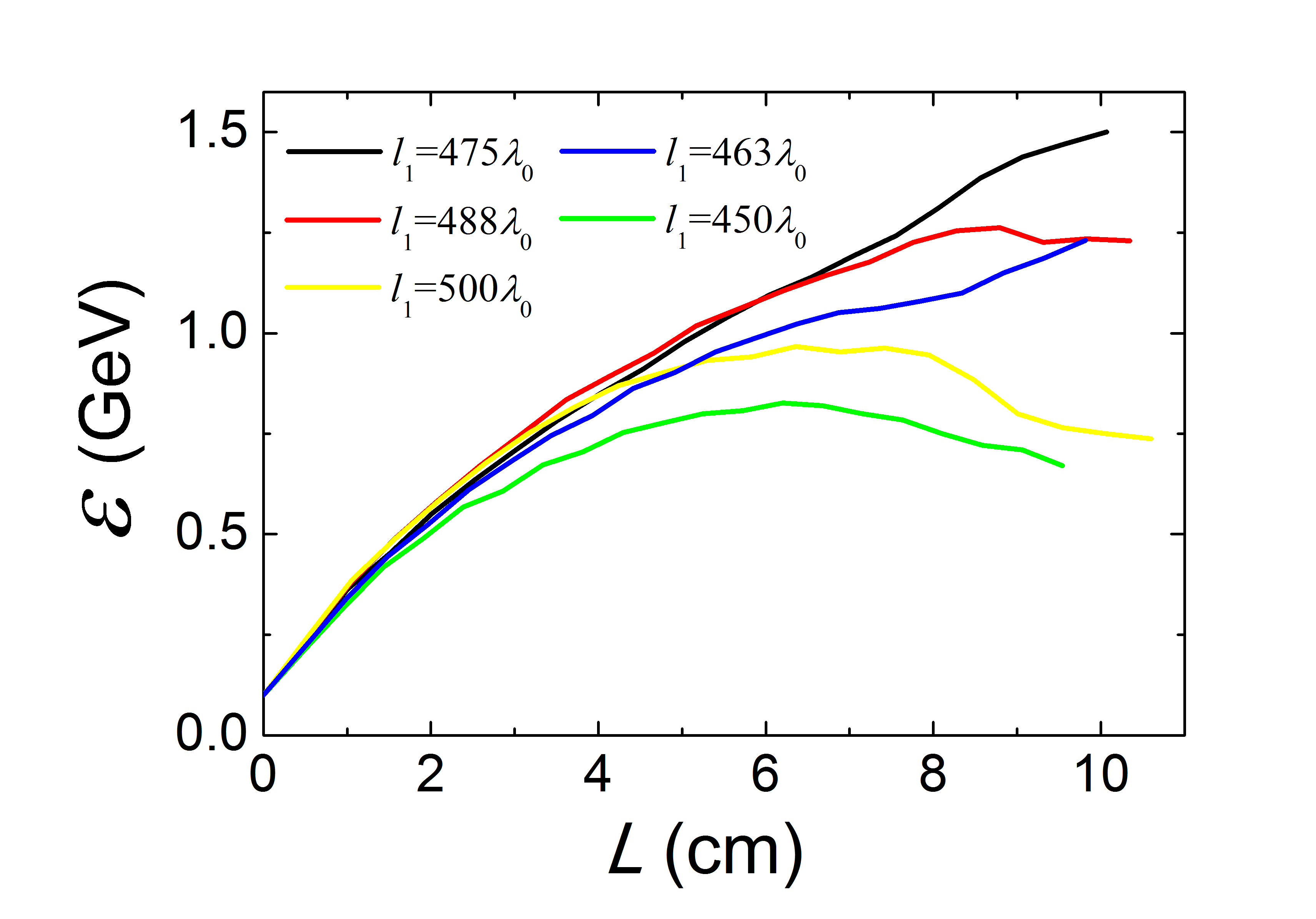}
\caption{(Color online.) Energy gain of injected electron bunch in CPW with different period.}
\label{fig:4}
\end{figure}

Equation~(10) indicates the dephasing length is very sensitive to
the transverse size of CPW, and it should be noted we have ignored
the small transverse motion of laser pulse for simplicity. According
to Eq.~(10), we know that the longitudinal dimension of CPW should
be chosen around the match condition $l_{1}=L_{d}\approx488\lambda_{0}$,
which roughly agrees with our numerical observations. By scanning
over a range of CPW periods, we found the optimal acceleration is
obtained for a slightly-shorter CPW with $l_{1}=475\lambda_{0}$ as
shown in Fig.~4. Apparently the violation of matching condition lead
to early saturation which limits the energy gain. The maximum energy
decreases 50$\%$ for 25 $\lambda_{0}$ deviation from the matched
cases.

In conclusion, a novel DLA scheme based on CPW at high laser intensities
is proposed and tested by multi-dimension PIC simulations. Our results
indicate that a CPW can be used as an electron accelerator when coupled
with state-of-art CO$_{2}$ laser beams. The proposed scheme demonstrates
high acceleration gradient and beam quality, which makes it a promising
candidate for future tabletop accelerator design. In addition, the
overall efficiency of the scheme is high because the CO$_{2}$ laser
pulses have high wall-plug efficiencies. The underlying physics is
discussed using PIC simulations and theoretical analysis, the matching
condition is presented, which agrees with our numerical observation.
The integration of longitudinal electric field in one CPW period results
in uniform accelerating structure in transverse direction. A quasi-monoenergetic
electron bunch with mean energy 1.5 GeV, r.m.s. energy spread 2$\%$ can
be obtained within 10-cm acceleration. Meanwhile, when a long electron
beam is injected into the CPW, micro-bunching effect comes into play,
which is capable of generating a chain of ultra-fast electron bunchs
with the dimension of half-laser-cycle.

\section*{Methods}
Due to the computational difficulty with simulating the realistic
long CO$_{2}$ laser pulse, we perform three dimensional (3D) simulation
with an relatively small window $L_{x}\times L_{y}\times L_{z}=12\lambda_{0}\times30\lambda_{0}\times20\lambda_{0}$
focused on the laser-electron interacting position over half of the
CPW period ($L_{acc}=475\lambda_{0}\approx0.5$ cm) to examine the
electron motion between two accelerating phases. The simulation resolution
is dx = 0.04 $\lambda_{0}$, dy = dz = 0.08 $\lambda_{0}$ in each direction,
where $\lambda_{0}=10.6\mu m$ is the wavelength of CO$_{2}$ laser.
In 2D simulations, a bigger simulation window ($L_{x}\times L_{y}=32\lambda_{0}\times26\lambda_{0}$)
and a finer resolution (dx = 0.02 $\lambda_{0}$, dy = 0.05 $\lambda_{0}$)
are employed, while other parameters remain the same. The laser pulse
in the window is assumed to have a trapezoidal profile in time
with normalized amplitude $a_{0}=5$, which propagates in the positive
$x$ direction. The plasma channel wall has a uniform density of $n=3n_{c}$, where
$n_{c}=m_{e}\omega_{0}^{2}/4\pi e^{2}$ is the critical density. It should be noted
although the density of MPW is limited by computational efficiency,
in real experiments, the laser can hardly penetrate into the area
with $n>3 n_{c}$ due to finite density gradients. In addition, the CPW contains a uniform
low-density ($10^{15}$ cm$^{-3}$) plasma to provide necessary focusing
for witness particles, it has little influence to the accelerating
field and laser propagation. The witness electron bunch has a flat
density profile in $x$ direction (duration $\approx4\lambda_{0}$)
and a Gaussian profile (FWHM $\approx2\lambda_{0}$) in transverse
direction.


\begin{thebibliography}{99}\suppressfloats
\bibitem{s1} {Tajima T. $\&$ Dawson, J.}, Laser electron accelerator. \emph{Phys. Rev. Lett.} \textbf{43}, 267 (1979).
\bibitem{s2} {Strickland D. $\&$ Mourou G.}, Compression of amplified chirped optical pulses. \emph{Opt. Commun.} \textbf{56}, 219 (1985).
\bibitem{bubble} {Pukhov A. $\&$ Meyer-ter-Vehn J.}, Laser wake field acceleration: the highly non-linear broken-wave regime. \emph{Appl. Phys. B} \textbf{74}, 355 (2002).
\bibitem{s3} {Mangles S. \emph{et al.}}, Monoenergetic beams of relativistic electrons from intense laser-plasma interactions. \emph{Nature} \textbf{431}, 535 (2004).
\bibitem{s4} {Faure J. \emph{et al.}}, A laser-plasma accelerator producing monoenergetic electron beams. \emph{Nature} \textbf{431}, 541 (2004).
\bibitem{s5} {Geddes C. \emph{et al.}}, High-quality electron beams from a laser wakefield accelerator using plasma-channel guiding. \emph{Nature} \textbf{431}, 538 (2004).
\bibitem{s6} {Leemans W. \emph{et al.}}, Multi-GeV electron beams from capillary-discharge-guided subpetawatt laser pulses in the self-trapping regime \emph{Phys. Rev. Lett.} \textbf{113}, 24502 (2014).
\bibitem{s7} {Pukhov A., Kostyukov I., Tuckmantel T., Luu-Thanh Ph., $\&$ Mourou G.}, Coherent acceleration by laser pulse echelons in periodic plasma structures. \emph{Eur. Phys. J. Special Topics} \textbf{223}, 1197 (2014).
\bibitem{s8} {Haberberger D., Tochitsky S., $\&$ Joshi C.}, Fifteen terawatt picosecond CO$_{2}$ laser system. \emph{Opt. Express} \textbf{18}, 17865 (2010).
\bibitem{s9} {Haberberger D. \emph{et al.}}, Collisionless shocks in laser-produced plasma generatemonoenergetic high-energy proton beams. \emph{Nat. Phys.} \textbf{8}, 95 (2012).
\bibitem{s10} {Pogorelsky I. \emph{et al.}}, Proton and ion beams generated with picosecond CO2 laser pulses. \emph{AIP Conf. Proc.} \textbf{1086}, 532 (2009).
\bibitem{s11} {Zhang L. \emph{et al.}}, High quality electron bunch generation with CO2-laser-plasma interaction \emph{Phys. Plasma} \textbf{22}, 023101 (2015).
\bibitem{s12} {Shen B., Zhao X., Yi L., Yu W., $\&$ Xu Z.}, Inertial confinement fusion driven by long wavelength electromagnetic pulses. \emph{High Power Laser Sci. Eng.} \textbf{1}, 105 (2013).
\bibitem{s13} {Lu W., Tzoufras M., Joshi C., Tsung F., $\&$ Mori W.}, Generating multi-GeV electron bunches using single stage laser wakefield acceleration in a 3D nonlinear regime. \emph{Phys. Rev. Spec. Top. - Accel. Beams} \textbf{10}, 061301 (2007).
\bibitem{s14} {Esarey E., Schroeder C., $\&$ Leemans W.}, Physics of laser-driven plasma-based electron accelerators. \emph{Rev. Mod. Phys.} \textbf{81}, 1229 (2009).
\bibitem{s15} {England, E. \emph{et al.}}, Dielectric laser accelerators. \emph{Rev. Mod. Phys.} \textbf{86}, 1337-1389 (2014).
\bibitem{s16} {Nanni E. \emph{et al.}}, Terahertz-driven linear electron acceleration.\emph{ Nat. Commun.} \textbf{6}, 1 (2015).
\bibitem{s17} {York A., Milchberg H., Palastro J., $\&$ Antonsen T.}, Direct acceleration of electrons in a corrugated plasma waveguide. \emph{Phys. Rev. Lett.} \textbf{100}, 195001 (2008).
\bibitem{s18} {Cros B. \emph{et al.}}, Eigenmodes for capillary tubes with dielectric walls and ultraintense laser pulse guiding. \emph{Phys. Rev. E} \textbf{65}, 026405 (2002).
\bibitem{s19} {Shen H.}, Plasma waveguide: A concept to transfer electromagnetic energy in space. \emph{J. Appl. Phys.} \textbf{69}, 6827 (1991).
\bibitem{s20} {Yi L., Pukhov A., Luu-Thanh Ph., Shen B.}, Bright X-ray source from a laser-driven microplasma waveguide. \emph{accepted} (2016).
\bibitem{s21} {Mizuno K., Pae J., Nozokido T., $\&$ Furuya K.}, Experimental evidence of the inverse Smith-Purcell effect. \emph{Nature} \textbf{328}, 45 (1987).
\bibitem{s22} {Breuer J., Graf R., Apolonski A., $\&$ Hommelhoff P.}, Dielectric laser acceleration of nonrelativistic electrons at a single fused silica grating structure: Experimental part. \emph{Phys. Rev. Spec. Top. - Accel. Beams} \textbf{17}, 021301 (2014).
\bibitem{s23} {Serafim P., Sprangle P., $\&$ Hafizi B.}, Optical guiding of a radially polarized laser beam for inverse cherenkov acceleration in a plasma channel. \emph{IEEE Trans. Plasma Sci.} \textbf{28}, 1155 (2000).
\bibitem{s24} {Jullien A., \emph{et al.}}, Highly efficient temporal cleaner for femtosecond pulses based on cross-polarized wave generation in a dual crystal scheme. \emph{Appl. Phys. B} \textbf{84}, 409 (2006).
\bibitem{s25} {Thaury C., \emph{et al.}}, Plasma mirrors for ultrahigh-intensity optics. \emph{Nat. Phys.} \textbf{3}, 424 (2007).
\bibitem{s26} {Fischer J. $\&$ Wegener M.}, Three-dimensional optical laser lithography beyond the diffraction limit. \emph{Laser Photonics Rev.} \textbf{7}, 22 (2013).
\bibitem{s27} {Cristoforetti G. \emph{et al.}}, Investigation on laser-plasma coupling in intense, ultrashort irradiation of a nanostructured silicon target \emph{Plasma Phys. Control. Fusion} \textbf{56}, 095001 (2014).
\bibitem{s28} {Kahaly S. \emph{et al.}}, Near-complete absorption of intense, ultrashort laser light by sub-$\lambda$ Gratings. \emph{Phys. Rev. Lett.} \textbf{101}, 145001 (2008).
\bibitem{s29} {Purvis M. \emph{et al.}}, Relativistic plasma nanophotonics for ultrahigh energy density physics. \emph{Nat. Photon.} \textbf{7}, 796 (2013).
\bibitem{s30} {Pukhov A.}, Three-dimensional electromagnetic relativistic particle-in-cell code VLPL. \emph{J. Plasma Phys.} \textbf{61}, 425 (1999).
\bibitem{s31} {Srinivasan R., Lobastov V., Ruan C., $\&$ Zewail A.}, Ultrafast electron diffraction (UED) a new development for the 4D determination of transient molecular structures. \emph{Helv. Chemi. Acta} \textbf{86}, 1763 (2003).
\bibitem{s32} {Dwyer J. \emph{et al.}}, Femtosecond electron diffraction: 'makingthe molecular movie' \emph{Phil. Trans. R. Soc. A} \textbf{364}, 741 (2006).
\bibitem{s33} {He Z. \emph{et al.}}, Electron diffraction using ultrafast electron bunches from a laser-wakefield accelerator at kHz repetition rate. \emph{Appl. Phys. Lett.} \textbf{102}, 064104 (2013).
\end{thebibliography}

\section*{Acknowledgements}
This work is supported by DFG Transregio TR18, EU FP7 EUCARD-2 projects and National Natural Science Foundation of China (No.11505262, No.11125526, and No.11335013).

\section*{Author contributions statement}
L. Q. Y. wrote the paper with contributions from A. P.; L. Q. Y. conducted simulations and analysis; A. P. developed the code used for simulations (VLPL) and supervised the work; B. F. S provided useful suggestions.

\section*{Additional information}
Competing financial interests: The authors declare no competing financial interests.

\end{document}